\documentclass[12pt]{article}
\usepackage{times}
\usepackage{geometry}
\usepackage[utf8]{inputenc}
\usepackage{enumitem,amssymb}
\usepackage{ragged2e}
\newlist{thematic}{itemize}{8}
\setlist[thematic]{label=$\square$}
\usepackage{pifont}

\usepackage{graphics,graphicx}
\begin{document}
\raggedright
\huge
Astro2020 Science White Paper \linebreak

Accretion in Stellar-Mass Black Holes at High X-ray Spectral Resolution \linebreak
\normalsize

\noindent \textbf{Thematic Areas:} \hspace*{60pt} $\square$ Planetary Systems \hspace*{10pt} $\square$ Star and Planet Formation \hspace*{20pt}\linebreak
$\boxtimes$ Formation and Evolution of Compact Objects \hspace*{31pt} $\square$ Cosmology and Fundamental Physics \linebreak
  $\square$  Stars and Stellar Evolution \hspace*{1pt} $\square$ Resolved Stellar Populations and their Environments \hspace*{40pt} \linebreak
  $\square$    Galaxy Evolution   \hspace*{45pt} $\square$             Multi-Messenger Astronomy and Astrophysics \hspace*{65pt} \linebreak
  
\textbf{Principal Author:}

Name:	Jon  M. Miller$^{1}$
 \linebreak						
Email: jonmm@umich.edu
 \linebreak

\textbf{Co-authors:} Didier Barret$^{2,3}$, Edward Cackett$^{4}$,
Maria Diaz Trigo$^{5}$, Christine Done$^{6}$, Elena Gallo$^{1}$, Jelle
Kaastra$^{7,8,9}$, Christian Motch$^{10}$, Ciro Pinto$^{11}$, Gabriele
Ponti$^{12,13}$, Natalie Webb$^{2}$, Abderahmen Zoghbi$^{1}$
\linebreak

\textbf{Abstract: Accretion disks around stellar-mass black holes
  offer unique opportunities to study the fundamental physics of
  standard thin disks, super-Eddington disks, and structure that may
  be connected to flux variability.  These local analogues of active
  galactic nuclei (AGN) are particularly attractive for their
  proximity, high flux, and peak emissivity in the X-ray band.  X-ray
  calorimeter spectrometers, with energy resolutions of 2-5 eV, are
  ideally suited to study accretion in stellar-mass black holes.  The
  results will make strong tests of seminal disk theory that applies
  in a broad range of circumstances, help to drive new numerical
  simulations, and will inform our understanding of AGN fueling,
  evolution, and feedback.}\\
\bigskip
\begin{footnotesize}
$^{1}$ Department of Astronomy, University of Michigan, 1085 South
  University Avenue, Ann Arbor, MI, 48103, USA\\
$^{2}$ IRAP CNRS, 9 Av. colonel Roche, BP 44346, F-31028
  Toulouse cedex 4, France\\
$^{3}$ Universite de Toulouse III Paul Sabatier / OMP,
  Toulouse, France\\
$^{4}$ Department of Physics \& Astronomy, Wayne State
  University, 666 West Hancock Street, Detroit, MI, 48201, USA\\
$^{5}$ ESO, Karl-Schwarzschild-Strasse 2, D-85748 Garching bei Munchen, Germany\\
$^{6}$ Department of Physics, University of Durham, South Road, Durham, DH1 3LE, UK\\
$^{7}$ SRON Netherlands Institute for Space Research,
  Sorbonnelaan 2, 3584 CA Utrecht, Netherlands\\
$^{8}$ Leiden Observatory, Leiden University, PO Box 2300 RA
  Leiden, Netherlands\\
$^{9}$ Department of Physics and Astronomy, Universiteit
  Utrecht, PO BOX 80000, 3508 TA Utrecht, Netherlands\\
$^{10}$ Universite de Strasbourg,CNRS, Observatoire Astronomique de
  Strasbourg, 11 rue de l'Universite, 67000, Strasbourg, France\\
$^{11}$ ESTEC/ESA, Keplerlaan 1, 2201AZ Noordwijk, The Netherlands\\
$^{12}$ INAF Osservatorio Astronomico di Brera, Via E. Bianchi 46,
  I-23807 Merate, Italy\\
$^{13}$ Max-Planck-Insitut fur Extraterrestrische Physik,
    Giessenbachstrasse, D-85748 Garching, Germany\\    
\end{footnotesize}  
\bigskip

\pagebreak

\noindent{\bf\large 1.0 Introduction}\\
\bigskip
Accretion disks are central to the growth and evolution of black
holes, and structure in the universe.  Mergers and accretion are both
likely to be important in black hole mass growth, but accretion may
dominate (Croton et al.\ 2006).  The {\it consequences} of accretion are
vast.  Disks around massive black holes have likely produced most of
the ionizing radiation in the universe since the epoch of reionization
(Peebles 2000, Fabian 2001).  The most extreme winds driven by
accretion disks may be able to seed the bulge of a host galaxy with
enough hot gas to halt star formation (e.g., Chartas et al.\ 2002,
Nardini et al.\ 2015).  The hot X-ray gas that dominates the baryonic
gas content of clusters can be reshaped by the jets driven by the
accretion disk around the massive black hole in the central galaxy
(Fabian 2012).\\
\bigskip
It is crucial, then, that we {\it observationally} test how accretion disks
work.  The fundamental analytical treatments of accretion disks are
now 40 years old (Shakura \& Sunyaev 1973, Blandford \& Payne 1982).
These seminal models predict that disk accretion is mediated by
magnetic processes; this has been verified and bolstered using more
recent numerical simulations (Balbus \& Hawley 1991, Miller \& Stone
2000).  Indeed, simulations have rapidly advanced with computing
power, and new efforts make detailed predictions of disk structure,
variability, winds, and jets (e.g., Ohsuga \& Mineshige 2011).\\
\bigskip
Disks around massive black holes peak in UV light, so measurements are
susceptible to neutral hydrogen scattering (both local to the source,
and spread within the host galaxy).  The long timescales natural to
massive black hole accretion further complicate the study of such
disks.  {\it Stellar-mass black holes may represent the optimal
laboratories for testing fundamental accretion disk physics.}  For a
broad range of Eddington fractions, disks around stellar-mass black
holes peak in X-rays, avoiding scattering by neutral hydrogen.  The
X-ray band is also fortuitous in that X-ray atomic lines are excited
by the X-ray continuum.  The ability to measure these simultaneously
readily facilitates studies of the physical processes and geometries
that define the accretion flow.\\
\bigskip
The High Energy Transmission Grating Spectrometer (HETGS; Canizares et
al.\ 2005) aboard {\it Chandra}, and the Reflection Grating
Spectrometer (RGS; Den Herder et al.\ 2001) aboard {\it XMM-Newton}
have demonstrated the ability of high-resolution X-ray spectroscopy to
reveal accretion flows (e.g., Boirin et al.\ 2005, Neilsen et
al.\ 2009).  In this white paper, we focus on a subset of particular
advances that will be possible using X-ray calorimeters, with
resolutions of 2--5 eV.  In the critical Fe K-shell band (roughly,
6-10 keV), these instruments will obtain spectra with $R = 1200-5000$,
comparable to ground- and space-based optical and UV spectrometers.\\
\bigskip
\bigskip

\noindent{\bf\large 2.0 The Inner Physics of Standard Accretion Disks}\\
\bigskip
Theory shows that disk accretion must be driven by magnetic processes:
internal viscosity via the magneto-rotational instability (or, MRI;
Balbus \& Hawley 1991) and/or magneto-centrifugal winds (Blandford \&
Payne 1982).  In fact, internal viscosity can also give rise to
magnetohydrodynamic winds (Proga 2003).  Although thermal continuum
emission from the disk {\it does not} bear the imprints of the processes
that created it, the line-rich outflows that result {\it do}.
High-resolution X-ray spectroscopy, then, can function as a window
into fundamental disk physics.
\bigskip

\begin{figure}[t!]
  \vspace{-0.5in}
  \includegraphics[scale=0.8]{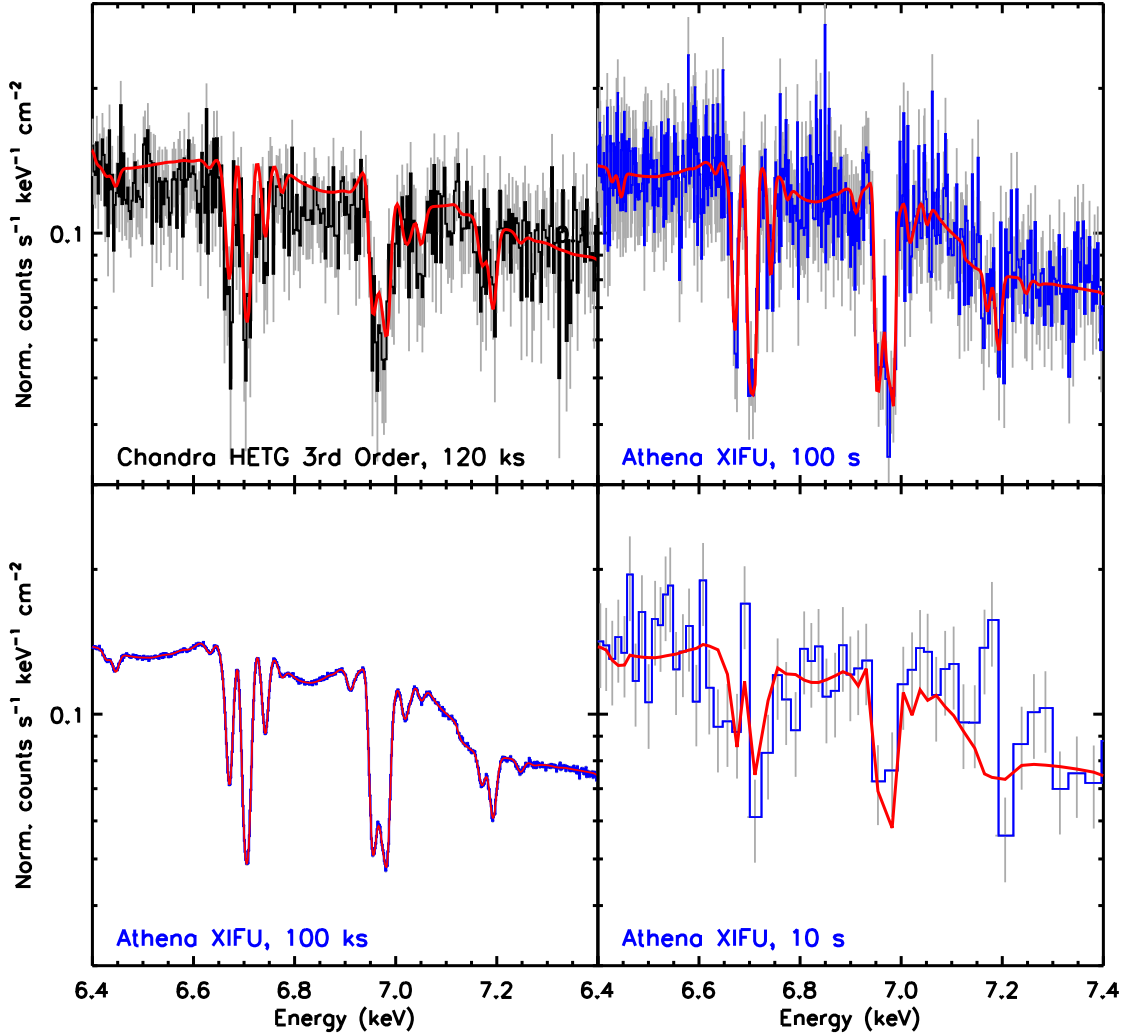}
\vspace{-0.25in}
\caption{\footnotesize The deep {\it Chandra}/HETG third-order spectrum of
  GRS 1915+105 (Miller et al.\ 2016) and simulated {\it Athena}/X-IFU
  (Barret et al.\ 2018) spectra constructed using the disk wind model
  for the {\it Chandra} data.  The complex at 6.7 keV is He-like Fe XXV
  (intercombination and resonance lines), the complex at 6.97 keV is
  He-like Fe XXVI (a spin-orbit doublet separated by 22 eV).  Features
  at 7.05 keV and 7.2 keV are Fe XXVI at an outflow velocities of
  0.01c of 0.03c.  Error bars are hardly visible in 100,000 seconds of
  X-IFU exposure (bottom left), and the level of detail achieved in
  100 seconds (top right) and 10 seconds (bottom right) exposures will
  enable studies of the wind on its {\it dynamical} time scales ($t\leq 200$
  seconds).}
\end{figure}

Interestingly, feedback between the inner accretion flow and outer
disk can influence this process.  Irradiation of the outer disk can
drive massive thermal winds that can potentially expel more gas than
is able to accrete inwards (Begelman et al.\ 1983); these flows are
also line-rich.  Heated gas that is unable to escape may supply the
ionized gas that gives rise to the non-thermal X-ray corona in black
hole systems (Shakura \& Sunyaev 1973).  Here again, high-resolution
X-ray spectroscopy is uniquely able to trace these processes.\\
\bigskip
Magnetohydrodymamic winds can be differentiated from
magnetocentrifugal winds using the run of ionization and density with
radius (e.g., the ``absorption measure distribution''; Behar 2009,
Fukumura et al.\ 2017).  The resolution of X-ray calorimeters will
make such measurements possible in a number of sources, facilitating
comparisons across the black hole mass scale.  The total mass outflow
rate and kinetic power in winds requires measurements of the gas
density and volume filling factor; the resolution afforded by
calorimeters will also make these determinations routine.  The
combination of high resolution, high throughput, and large collecting
area possible with, e.g., the {\it Athena} X-ray Integral Field Unit (or
X-IFU; Barret et al.\ 2018), for instance, will enable transformative
studies winds on their dynamical timescales ($t \propto r/v$, where $r$
is the wind launching radius and $v$ is the outflow velocity).  This
will provide an entirely different angle on wind launching mechanisms
via velocity variations, density variations, and ionization changes,
and will thereby enable quantitatively and qualitatively new
comparisons to detailed numerical simulations of accretion disks.
{\bf Figure 1} shows simulated {\it Athena}/X-IFU spectra of the stellar-mass
black hole GRS 1915+105 on the {\it dynamical} timescales of its disk wind,
based on a recent deep {\it Chandra}/HETGS observation (Miller et
al.\ 2016).\\
\bigskip

\begin{figure}[t!]
  \vspace{-0.5in}
  \includegraphics[scale=0.64,angle=-90]{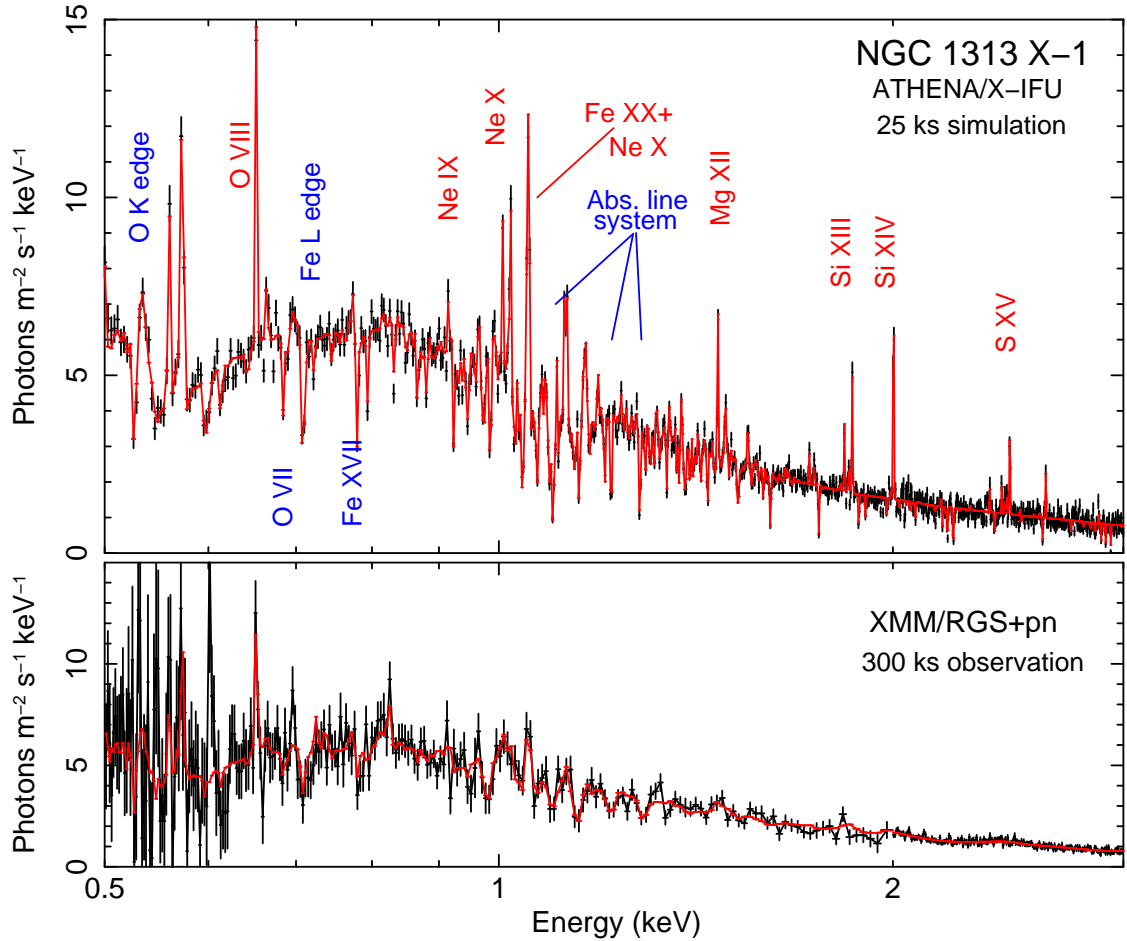}
\vspace{0.2in}
\caption{\footnotesize A simulated 25 ks Athena/X-IFU exposure of NGC
  1313 X-1, a nearby ULX that may harbor a stellar-mass black hole
  accreting well above the Eddington limit. Extremely deep XMM-Newton
  RGS spectra suggest a combination of emission and absorption lines
  consistent with a $v=0.2c$ outflow.  In just 25 ks of exposure,
  Athena/X-IFU spectra will reveal fast outflows characteristic of
  super-Eddington accretion, and examine the variability of such
  flows.  This will be possible in a sample of nearby ULXs, enabling
  population studies of the source class.  The gratings spectrometers
  envisioned for missions like {\it ARCUS} and {\it Lynx} would also
  excel in studies of ULX outflows.}
\end{figure}

\bigskip
\noindent{\bf\large 3.0 Super-Eddington Accretion Disks}\\
\bigskip
Luminous quasars are evident in the early universe (e.g., Fan et
al.\ 2001), implying that some black holes were quickly able to reach
masses exceeding a billion solar masses.  New detections of quasars at
$z \geq 7$ pose even more severe problems: a stellar-mass black hole
accreting continuously at the Eddington limit will not reach such
masses at the redshifts they are observed (Mortlock et al.\ 2011,
Banados et al.\ 2018).  Thus, although mergers may be important, it is
likely that super-Eddington accretion played a key role (Volonteri et
al.\ 2006, 2016).  Understanding super-Eddington accretion is
therefore fundamental to understanding the evolution of the universe.
However, surveys of the low- and moderate- redshift universe reveal
that both quasars and super-Eddington accretion are rare (e.g, Hickox
et al.\ 2009).  Therefore, we must turn to other settings to
understand this critical process.  Ultra-luminous X-ray sources (ULXs)
are bright, off-nuclear sources that exceed the Eddington limit for 10
solar-mass black holes (Fabbiano et al.\ 1989).  A set of ULXs are
known to be pulsars (Bachetti et al.\ 2014), but other sources may be
stellar-mass black holes accreting far above the Eddington limit.\\
\bigskip
Summing several {\it XMM-Newton}/RGS observations of ULXs NGC 1313 X-1 and
NGC 5408 X-1, emission and absorption lines are finally detected
(Pinto et al.\ 2016).  The fastest absorbers indicate velocities of $v
= 0.2c$, suggesting that radiation is coupling to moderately ionized
gas, as per radiation driving of a super-Eddington wind.
Intriguingly, fast winds in these systems - and so-called “ultra-fast
outflows” (or, UFOs) in AGN - appear to be transient (Reeves et
al.\ 2018, Pinto et al.\ 2018).  This may be consistent with a
radiation force multiplier effect that only operates within a narrow
range of gas ionization (Proga 2003).\\
\bigskip
The resolution of X-ray calorimeter spectrometers is fixed in energy
space, so they deliver higher resolution at higher energy.  However,
they can still deliver revolutionary spectra at low energy, aided by
the fact that they are not dispersive spectrometers.  This efficiency
gain is best leveraged by a large collecting area and high throughput.
{\bf Figure 2} shows a simulated 25 ks {\it Athena}/X-IFU spectrum of NGC 1313
X-1, based on the outflow observed in {\it XMM-Newton}/RGS data (Pinto et
al.\ 2016).  In just 25 ks of exposure, numerous emission and
absorption lines can be detected, and the total mass outflow rate and
kinetic power of the wind can be measured.  The ability to study
super-Eddington outflows on short timescales will reveal the physical
processes that make them transient.  Comparing ULX spectra to SS 433
in the Milky Way may reveal additional disk accretion and ejection
physics.\\
\bigskip

\begin{figure}[t!]
\hspace{1.0in}
  \includegraphics[scale=0.6]{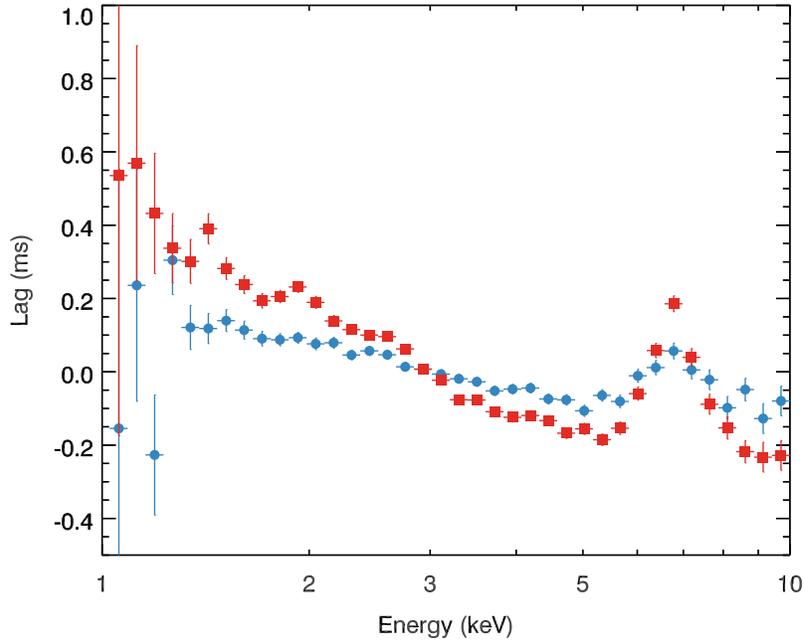}
\vspace{-0.25in}
\caption{\footnotesize Simulated {\it Athena}/X-IFU lag spectra of an
  accretion disk with (red squares) and without (blue circles)
  heightened reflection from $r = 100-300~ GM/c^{2}$.  This simulated
  exposure is just 4 ks; the dual-reflector model recently considered
  for MAXI J1535-571 was assumed (Miller et al.\ 2018).  The lag
  spectra were generated using the methods described in Cackett et
  al. (2014).  The combination of time-averaged spectra and
  reverberation spectroscopy that is possible with a sensitive, high
  throughput calorimeter can clearly reveal disk warps and tearing.
  Similar results would be possible with an instrument like the {\it Athena}
  Wide-Field Imager.}
\vspace{-0.15in}
\end{figure}

\noindent{\bf\large 4.0 Revealing Disk Structure}\\
\bigskip
Standard disks are expected to be extremely thin close to the
innermost stable circular orbit (or, ISCO), and the disk scale height
is expected to increase smoothly at larger radii (Shakura \& Sunyaev
1973).  {\it However, it is possible that actual disks behave
  differently.}  Radiation pressure on larger radii from dissipation
close to the ISCO may give rise to warps that change the local
contours of the disk (Maloney et al.\ 1996).  In transitional states,
wherein the accretion flow changes from disk-dominated to
corona-dominated, and wherein the outflow mode may be changing from
wind- to jet-dominated (Gallo et al.\ 2003), disk “tearing” may cause
some annuli to be disconnected and lie out of the broader disk plane
(Nixon \& Salvesen 2015).  Some quasi-periodic oscillations (QPOs) may
be due to structures such as warps, and potentially connected to
Lense-Thirring precession (Miller \& Homan 2005, Ingram et
al.\ 2016).\\
\bigskip
X-ray irradiation of a disk with a smoothly varying scale height gives
rise to a standard disk reflection spectrum, most prominently revealed
through Fe K emission lines (e.g., Miller 2007).  However, if specific
annuli have anomalous scale heights - as might be the case if the disk
is warped or “torn” - this can be revealed through disk reflection.
X-ray calorimeters offer unsurpassed spectral resolution, but they can
also function as excellent instruments for X-ray timing.  If
structures like warps or tearing are manifest in stellar-mass black
hole disks, they will be detected with an instrument like the {\it
  Athena}/X-IFU through time-averaged spectroscopy, and - more
importantly - through reverberation lag spectra.  Observations with
spectrometers with moderate resolution but excellent time resolution,
such as the {\it Athena} Wide-Field Imager (or, WFI; Meidinger et
al.\ 2017), can also capture this science.  {\bf Figure 3} shows
simulated reverberation lag spectra from a short (4 ks) observation of
a bright transient source like GX 339-4, using the dual-reflector
model required in recent NICER observations of MAXI J1535-571 (Miller
et al.\ 2018).  Following Cackett et al. (2014), reverberation can
easily be detected from radii as small as 6 GM/c2 but longer lags
accumulate from structure at $r = 100-300 GM/c^{2}$.  The combination
of time-averaged and lag spectroscopy afforded by X-ray calorimeters
will be a powerful means of testing disk structures, and potentially
the origins of X-ray QPOs.\\


\clearpage
\textbf{References}\\
Bachetti, M., et al., 2014, Nature, 514, 202.\\
Balbus, S., \& Hawley, J., 1991, ApJ, 376, 214.\\
Banados, E., et al., 2018, ApJ, 861, L14.\\
Barret, D., et al., 2018, SPIE, 10699.\\
Begelman, M., et al., 1983, ApJ, 271, 70.\\
Behar, E., 2009, ApJ, 703, 1346.\\
Blandford, R., \& Payne, D., 1982, MNRAS, 199, 883.\\
Boirin, L., et al., 2005, A\&A, 436, 195.\\
Cackett, E., et al., 2014, MNRAS, 438, 2980.\\
Canizares, C., et al., 2005, PASP, 117, 1144.\\
Chartas, G., et al., 2002, ApJ, 579, 169.\\
Croton, D., et al., 2006, MNRAS, 365, 11.\\
Den Herder, et al., 2001, A\&A, 365, L7.\\
Fabbiano, G., 1989, ARA\&A, 27, 87.\\
Fabian, A., 2001, AIPC, 599, 93.\\
Fabian, A., 2012, ARA\&A, 50, 455.\\
Fan, X., et al., 2001, AJ, 121, 54.\\
Fukumura, K., et al., 2017, Nature Astronomy, 1, 62.\\
Gallo, E., et al., 2003, MNRAS, 344, 60.\\
Hickox, R., et al., 2009, ApJ, 696, 891.\\
Ingram, A., et al., 2016, MNRAS, 461, 1967.\\
Maloney, P., et al., 1996, 472, 582.\\
Meidinger, N., et al., 2017, SPIE, 10297, 0.\\
Miller, J. M., 2007, ARA\&A, 45, 441.\\
Miller, J. M, \& Homan, J., 2005, ApJ, 618, L107.\\
Miller, J. M., et al., 2016, ApJ, 821, L9.\\
Miller, J. M., et al., 2018, ApJ, 860, L28.\\
Miller, K., \& Stone, J., 2000, ApJ, 534, 398.\
Mortlock, D. J., et al., 2011, Nature, 474, 616.\\
Nardini, E., et al., 2015, Science, 347, 860.\\
Neilsen, J., Lee, J., 2009, Nature, 458, 481.\\
Nixon, C., \& Salvesen, G., 2014, MNRAS, 437, 3994.\\
Ohsuga, K., \& Mineshige, S., 2011, ApJ, 736 2.\\
Peebles, J., 2000, astro-ph/0010617.\\
Pinto, C., et al., 2016, Nature, 533, 64.\\
Pinto, C., et al., 2018, MNRAS, 476, 1021.\\
Proga, D., 2003, ApJ, 585, 406.\\
Reeves, J., et al., 2018, ApJ, 854, 28.\\
Shakura, N., \& Sunyaev, R., 1973, A\&A, 24, 337.\\
Volonteri, M., et al., 2006, ApJ, 650, 669.\\
Volonteri, M., et al., 2016, MNRAS, 460, 2979.\\

\end{document}